\documentclass[11pt,doublespace]{article}
\usepackage{graphicx}
\usepackage{epsfig}
\usepackage{amsmath}
\usepackage{amssymb}

\newcommand \be{\begin{equation}}
\newcommand \ba{\begin{eqnarray}}
\newcommand \ee{\end{equation}}
\newcommand \ea{\end{eqnarray}}

\begin{document}

\begin{center}
{\LARGE Predictability of large future changes in major financial indices}
\end{center}
\bigskip
\begin{center}
{\large Didier
Sornette{\small$^{\mbox{\ref{igpp},\ref{ess},\ref{lpec}}}$},
Wei-Xing Zhou{\small$^{\mbox{\ref{igpp}}}$}}
\end{center}
\bigskip
\begin{enumerate}
\item Institute of Geophysics and Planetary Physics, University of
California, Los Angeles, CA 90095\label{igpp} \item Department of
Earth and Space Sciences, University of California, Los Angeles,
CA 90095\label{ess} \item Laboratoire de Physique de la Mati\`ere
Condens\'ee, CNRS UMR 6622 and Universit\'e de Nice-Sophia
Antipolis, 06108 Nice Cedex 2, France\label{lpec}
\end{enumerate}

Correspondence should be address to D. SORNETTE:

1693 Geology Building\\
Institute of Geophysics and Planetary Physics\\
University of California, Los Angeles\\ CA 90095-1567, USA.

Tel: +1-310-825-2863\\ Fax: +1-310-206-3051.\\ {\it E-mail
address:}\/sornette@moho.ess.ucla.edu (D. Sornette)

\newpage

\bigskip
\bigskip
\bigskip
\bigskip
\bigskip
\bigskip
\bigskip
\bigskip
\bigskip
\bigskip
\bigskip
\bigskip
\bigskip
\bigskip
\bigskip
\bigskip
\bigskip
\bigskip
\bigskip
\bigskip
\bigskip
\bigskip
\bigskip
\bigskip
\bigskip
\bigskip

--------------------------
\bigskip
\bigskip
\bigskip
\bigskip
\bigskip
\bigskip
\bigskip
\bigskip
\bigskip
\bigskip
\bigskip
\bigskip
\bigskip
\bigskip
\bigskip
\bigskip
\bigskip
\bigskip

\begin{center}
{\LARGE Predictability of large future changes in major financial indices}
\end{center}

\newpage

\begin{abstract}
We present a systematic algorithm testing for the existence of
collective self-organization in the behavior of agents in social
systems, with a concrete empirical implementation on the Dow Jones
Industrial Average index (DJIA) over the 20th century and on Hong
Kong Hang Seng composite index (HSI) since 1969. The algorithm
combines ideas from critical phenomena, the impact of agents'
expectation, multi-scale analysis and the mathematical method of
pattern recognition of sparse data. Trained on the three major
crashes in DJIA of the century, our algorithm exhibits a
remarkable ability for generalization and detects in advance 8
other significant drops or changes of regimes. An application to
HSI gives promising results as well. The results are robust with
respect to the variations of the recognition algorithm. We
quantify the prediction procedure with error diagrams.
\end{abstract}

{\it{Keywords}}: Econophysics; Multiscale analysis; Pattern
recognition; Predictability

\newpage

\section{Introduction}

It is widely believed that most complex systems are unpredictable,
with concrete implications in earthquake prediction\footnote{See
the contributions in Nature debates on earthquake prediction at
http://helix.nature.com/debates/earthquake/.}, in engineering
failure \cite{Karplus}, and in financial markets
\cite{famaefficient} to cite a few. In addition to the persistent
failures of predictive schemes for these systems, concepts such as
self-organized criticality \cite{SOC} suggest an intrinsic
impossibility for the prediction of catastrophes. Several recent
works suggest a different picture: catastrophes may result from
novel mechanisms amplifying their size \cite{S99a,S99b,S99c} and
may thus be less unpredictable than previously thought. This idea
has been mostly explored in material failure
\cite{critrup1,critrup2,critrup3}, in earthquakes \cite{KBbook},
and in financial markets and emerged in the latter from the
analysis of cumulative losses (drawdowns) \cite{dd1,dd2}, from
measures of algorithmic complexity \cite{Mansilla}, and from
agent-based models \cite{Johnson}.

We present novel empirical tests that provide a strong support for
the hypothesis that large events can be predicted.
We focus our analysis on financial indices
(typically the daily Dow Jones Industrial Average (DJIA) from
26-May-1896 to 11-Mar-2003) as they provide perhaps the best data
sets that can be taken as proxies for other complex systems. Our
methodology is based on the assumption that fast large market
drops (crashes) are the results of interactions of market players
resulting in herding behavior: exogenous shocks (due to changes in
market fundamentals) often do not play an important role, which is
at odds with standard economy theory. The fact that exogenous
shocks may not be the most important driving causes of the
structures found in financial time series has been shown to be the
case for volatility shocks \cite{endoexo}: most of the bursts of
volatility in major US indices can be explained by an endogenous
organization involving long-memory processes, while only a few
major shocks such as 9/11/2001 or the coup against Gorbachev in
1991 are found to leave observable signatures. Concerning
financial crashes in indices, bonds and currencies, \cite{JS03a}
have performed an extended analysis of the distribution of
drawdowns (cumulative losses) in the two leading exchange markets
(US dollar against the Deutsch and against the Yen), in the major
world stock markets, in the U.S. and Japanese bond market and in
the gold market and have shown the existence of ``outliers,'' in
the sense that the few largest drawdowns do not belong to the same
distribution. For each identified outlier, \cite{JS03a} have
checked whether the formula (\ref{Eq:lnpt}) given below,
expressing a so-called log-periodic power law signature (LPPL),
could fit the price time series preceding them; if yes, the
existence of the LPPL was taken as the qualifying signature for an
endogenous crash (a drawdown outlier was seen as the end of a
speculative unsustainable accelerating bubble generated
endogenously). In the absence of LPPL, \cite{JS03a} were able to
identify the relevant historical event, {\it i.e.}, a new piece of
information of such magnitude and impact that it is reasonable to
attribute the crash to it, following the standard view of the
efficient market hypothesis. Such drawdown outliers were thus
classified as having an exogenous origin. Globally over all the
markets analyzed, \cite{JS03a} identified 49 outliers, of which 25
were classified as endogenous, 22 as exogenous and 2 as associated
with the Japanese ``anti-bubble'' starting in Jan. 1990.
Restricting to the world market indices, \cite{JS03a} found 31
outliers, of which 19 are endogenous, 10 are exogenous and 2 are
associated with the Japanese anti-bubble. The combination of the
two proposed detection techniques, one for outliers in the
distribution of drawdowns and the second one for LPPL, provided a
systematic taxonomy of crashes. The present paper goes one step
further to propose a prediction scheme for crashes, and by
extension large and rapid changes of regime.

Our proposed learning algorithm belongs to a recent body of
literature that challenges the Efficient Market Hypothesis (EMH).
In the context of Information Theory which is related to the
present pattern recognition approach, see for instance
\cite{Shmilovici}. Our approach is also related to the field of
nonlinearity applied to financial time series: see
\cite{Brockreview} for a review and
\cite{Hsieh1,Brock2,Hsieh2,Kaboudan,BrockHommes1,Harriff,BrockHommes2}
for tests and models of nonlinearity in financial time series.
However, while we use a deterministic function to fit bubble
periods, our approach is not deterministic because our underlying
rational expectation bubble theory acknowledges the stochastic
component introduced by the crash hazard rate and by the random
regime shifts associated with the nucleation of bubble periods. In
addition, the LPPL pattern that we propose to characterize these
bubble periods, while rather rigid in its geometrical proportions
exemplifying the concept of discrete scale symmetry (see below),
is different from one bubble to the next as its time span and
amplitude can change.

This paper is organized as follows. Section \ref{s1:Theory}
presents the theoretical foundation of our approach. Section
\ref{s1:PR} describes the definition of the pattern recognition
method and its implementation and tests on the DJIA. Section
\ref{s1:HSI} shows further tests on the Hong Kong Hang-Seng index
and section \ref{s1:conclude} concludes.

\section{Theoretical foundation of our approach}
\label{s1:Theory}

Our key idea is to test for signatures of collective behaviors
similar to those well-known in condensed-matter and statistical
physics. The existence of similar collective behaviors in
physics and in markets may be surprising to those
not involved in the study of complex systems. Indeed, in physics, the governing
laws are well established and tested whereas one could argue
that there is no well-established
fundamental law in stock markets. However, the science of complexity
developed in the last two decades has shown that the emergence of
collective phenomena proceeds similarly in physics, biology or
social sciences as long as one is interested in the coarse-grained
properties of large assemblies of constituting elements or agents
(see for instance \cite{Anderson,SOC,Review1,crit_Sor} and references therein).

Collective behaviors in population of agents can emerge through
the forces of imitation leading to herding
\cite{herding1,herding2,herding3}. Herding behavior of investors
is reflected in significant deviations of financial prices from
their fundamental values \cite{Shillerexu}, leading to so-called
speculative bubbles \cite{Flood} and excess volatility
\cite{Shillervol}. The similarity between herding and statistical
physics models, such as the Ising model, has been noted by many
authors
\cite{Caleenmon1,Caleenmon2,Caleenmon3,Caleenmon4,Review1,Review2}.
We use this similarity to model a speculative bubble as resulting from positive
feedback investing, leading to a faster-than-exponential power law
growth of the price \cite{JSL1,JSL2,JSL3}. In addition, the
competition between such nonlinear positive feedbacks, negative
feedbacks due to investors trading on fundamental value and the
inertia of decision processes lead to nonlinear oscillations
(nonlinear in the sense of amplitude-dependence frequencies)
\cite{idesor} that can be captured approximately by adding an
imaginary part to the exponent of the power law price growth (by
power law price growth, we refer to the super-exponential growth
typically associated with the solution of the equation $dp/dt =
r(p) p$, with $r(p) \sim p^\delta$ and $\delta>0$. For $\delta=0$,
one recovers the standard exponential growth. For $\delta>0$,
$p(t)$ can approach a singularity in finite time like $p(t) \sim
|t_c-t|^{\alpha}$; its discrete time approximation takes the form
of an exponential of an exponential of time \cite{expexp}). The
addition of an imaginary part to the exponent $\alpha$ such that
it is written $\alpha=\alpha'+i\alpha''$ describes the existence
of accelerating log-periodic oscillations obeying a discrete scale
invariance (DSI) symmetry \cite{dsi,Review1,Review2}. This can be
seen from the fact that $|t_c-t|^{\alpha'+i\alpha''}=
|t_c-t|^{\alpha'}\cdot \exp[i\alpha'' \ln |t_c-t|]$ whose real
part is $|t_c-t|^{\alpha'}\cdot \cos[\alpha'' \ln |t_c-t|]$. More
generally, it can be shown that the first terms in a systematic
expansion can be written as
\begin{equation}
\ln[\pi(t)] = A - B\tau^m + C\tau^m
\cos\left[\omega\ln(\tau)-\phi\right]~, \label{Eq:lnpt}
\end{equation}
where $\pi(t)$ is the price of a specific asset (it can be a
specific stock, commodity or index), $A$ is the logarithm of the
price at $t_c$, $\omega$ is the angular log-frequency, $B>0$ and
$0<m<1$ for an acceleration to exist, $C$ quantifies the amplitude
of the log-periodic oscillations, $\phi$ is an arbitrary phase
determining the unit of the time, $t$ is the current time, and
$\tau=t_c-t$ is the distance to the critical time $t_c$ defined as
the end of the bubble. Equation (\ref{Eq:lnpt}) does not hold for
any time but rather applies to those non-stationary transient
phases, otherwise known as bubbles.

Expression (\ref{Eq:lnpt})
has been used to fit bubbles according to the following algorithm described in
\cite{emergent}, that we summarize here: (i) the crashes were identified
within a given financial time series as the largest cumulative losses
(largest drawdowns) according to the methodology of \cite{dd2}; (ii) the
bubble preceding a given crash was then identified as the price trajectory
starting at the most obvious well-defined minimum before a ramp-up
ending at the highest value before the crash. The starting value
(local minimum)
of the bubble is unambiguous for large developed markets but it was sometimes
necessary to change the interval fitted due to
some ambiguity with respect to the choice of the first point in
these smaller emergent markets \cite{dd2}. This algorithm was applied
to over thirty crashes in the major financial markets and the fits
of their corresponding bubbles with Eqn.(\ref{Eq:lnpt}) showed
that the DSI parameter $\omega$ exhibits a remarkable
universality, with a mean of $6.4$ and standard deviation of $1.6$
\cite{JS03a,JS03b}, suggesting that it is a significant
distinguishing feature of bubble regimes. In these previous look back studies,
the knowledge of the time of a given crash was used to identify its
associated bubble with the purpose of testing for a common behavior. This was
done in the usual spirit of a scientific endeavor which consists
in general in the following steps: (a) collect samples;
(b) test for regularities and commonalities in the samples to establish a
robust classification of a new phenomenon;
(c) develop a model accounting for this classification and calibrate it;
(d) put the model to test using its prediction that are applied to novel data,
that have not be used in the model construction. The previous works
mentioned above \cite{JS03a,JS03b} cover the steps (a-c). The present paper
addresses step (d) by developing a systematic look forward approach in which
the goal is to identify ``in real time'' the existence of a
developing bubble and
to predict times of increased crash probability.

The physics-type construction (i.e.,
emphasizing interactions and collective behavior)
leading to Eqn.(\ref{Eq:lnpt}) is however missing an essential
ingredient distinguishing the natural and physical from the social
sciences, namely the existence of expectations: market
participants are trying to discount a future that is itself shaped
by market expectations. As exemplified by the famous parable of
Keynes' beauty contest, in order to predict the winner,
recognizing objective beauty is not very important, but knowledge
or prediction of others' prediction of beauty is much more
relevant. Similarly, mass psychology and investors' expectations
influence financial markets significantly. We use the
rational-expectation (RE) Blanchard-Watson model of speculative
bubbles \cite{RE1,RE2} to describe concisely such feedback loops
of prices on agent expectation and vice-versa. Applied to a given
bubble price trajectory, agents form an expectation of the
sustainability of such bubble and of its potential burst. By this
process, to each price $\pi(t)$ is associated a crash hazard rate
$h(t)$, such that $h(t) dt$ is the probability for a crash or
large change of regime of average amplitude $\kappa$ to occur
between $t$ and $t+{d}t$ conditioned on the fact that it has not
happened. The crash hazard rate $h(t)$ quantifies the general belief
of agents in the
non-sustainability of the bubble. The RE-bubble model predicts the
remarkable relationship \cite{JSL1,JSL2,JSL3}
\begin{equation}
\ln\left[\frac{\pi(t)}{\pi(t_0)}\right] = \kappa\int_{t_0}^t  h(t')dt'
\label{positcond}
\end{equation}
linking price to crash hazard rate. This equation is the solution of
the model describing the
dynamics of the price increment $d\pi$ over
a time increment dt: $d\pi = \mu \pi dt - \kappa \pi dj$, where $\mu$
is the instantaneous return,
$dj$ is a jump process equal to $1$ when a crash occurs with amplitude $\kappa$
and zero otherwise. The no-arbitrage condition ${\rm E}[d\pi]=0$ implies that
$\mu \pi = \kappa \pi {\rm E}[dj/dt]$, where ${\rm E}[.]$ denotes the
expectation
conditional on all information available up to the present time. By
definition of the crash hazard rate, $h(t) = {\rm E}[dj/dt]$, this gives
$\mu \pi = \kappa \pi h(t)$. Thus, conditioned on no crash having
occurred until time $t$,
the integration of the model yields (\ref{positcond}).

Substituting Eqn.(\ref{Eq:lnpt}) in
Eqn.(\ref{positcond}) and using the fact, that by definition $h(t)
\geq 0$, gives the constraint
\begin{equation}
b \equiv B m - |C| {\sqrt{m^2 + \omega^2}} \geq 0~.
\label{defb}
\end{equation}
This condition has been found to provide a
significant feature for distinguishing market phases with bubbles
ending with crashes from regimes without bubbles \cite{BC02PhysA}.

Note that $h(t)$ is not normalized to $1$: $\int_0^{+\infty} h(t)
dt$ gives the total conditional probability for a crash to occur,
which is less than $1$ for the RE bubble model to hold. There is
thus a finite probability $1-\int_0^{+\infty} h(t) dt$ for the
bubble to land smoothly without a crash. This is necessary for the
RE bubble model, as otherwise the crash would be certain. By
having a non-zero probability for no crash to occur, it remains
rational for investors to remain in the market and harvest the
risk premium (the bubble) for the risk of a possible (but not
certain) crash. Technically, using the standard rule of
conditional probabilities, $h(t)$ is related to the unconditional
probability $p(t)$ for a crash per unit time by $h(t) =
p(t)/\int_t^{+\infty} p(t') dt'$. By integrating, rearranging and
differentiating, this gives $p(t) = h(t) \exp[-\int_0^t h(t')
dt']$.

\section{The pattern recognition method}
\label{s1:PR}

\subsection{Objects and classes}

Let us now build on these insights and construct a systematic
predictor of bubble regimes and their associated large changes.
Due to the complexity of financial time series, to the existence
of many different regimes, and to the constant action of investors
arbitraging gain opportunities, it is widely held that the
prediction of crashes is an impossible task \footnote{A.
Greenspan, Economic Volatility, Remarks at symposium sponsored by
the Federal Reserve Bank of Kansas City, Jackson Hole, Wyoming,
August 30, 2002.}. The truth is that financial markets do not
behave like statistically stationary physical systems and exhibit
a variety of non-stationarities, influenced both by exogenous and
endogenous shocks \cite{endoexo}. To obtain robustness and to
extract information systematically in the approximately
self-similar price series, it is useful to use a multiscale
analysis: the calibrations of Eqn.(\ref{Eq:lnpt}) and
Eqn.(\ref{positcond}) are thus
performed for each trading day over nine different time scales of
$60$, $120$, $240$, $480$, $720$, $960$, $1200$, $1440$ and $1680$
trading days.

Each trading day ending an interval of length $1680$ or smaller is
called an object and there are a total of 27428 objects from
06-Feb-1902 till 11-Mar-2003.
We define a map from the calendar
time $t$ to an ordered index $\$t$ of the trading day, such that
$\$t=t$ if $t$ is a trading date and $\$t=\emptyset$ (empty set) otherwise.
This mapping is just a convenient way for defining the set of objects
to which our method is applied. It does not produce any bias or
additional dynamics since it is used only in the classification scheme.
We denote
${\mathcal{T_C}} = \{t_{c,i}: i=1,\cdots,n\}$ a list of targets
(crashes) that happened at $t_{c,i}$. The total number of targets is thus $n$.
All objects are partitioned
into two classes ${\rm{I}}$ and ${\rm{II}}$, where class
${\rm{I}}(t_l) = \{\$t: t_{c,i}-t_l \le t\le t_{c,i}, t_{c,i} \in
{\mathcal{T_C}} \}$ contains all the objects in $[t_{c,i}-t_l, \le
t_{c,i}]$ for $i=1,\cdots,n$ and class ${\mathcal\rm{II}}(t_l) =
\{T: T=1,\cdots,27428\} - {\rm{I}}$ includes all the remaining,
where $t_l$ is a given number in unit of calender day. In the
following, our target set are three well-known speculative bubbles
that culminated on 14-Sep-1929, 10-Oct-1987 and 18-Jul-1998,
respectively, before crashing in the following week or month.
The first two bubbles are perhaps the most famous or infamous examples
discussed in the literature. The last one is more recent and has been
associated with the crisis of the Russian default on its debt and LTCM.
Of course, the choice of this triplet is otherwise arbitrary. Our goal
is just to show that our method can generalize to other bubbles beyond
those which have been used for training.

\subsection{Empirical tests of the relevance of patterns}

\begin{figure}
\includegraphics[width=13cm]{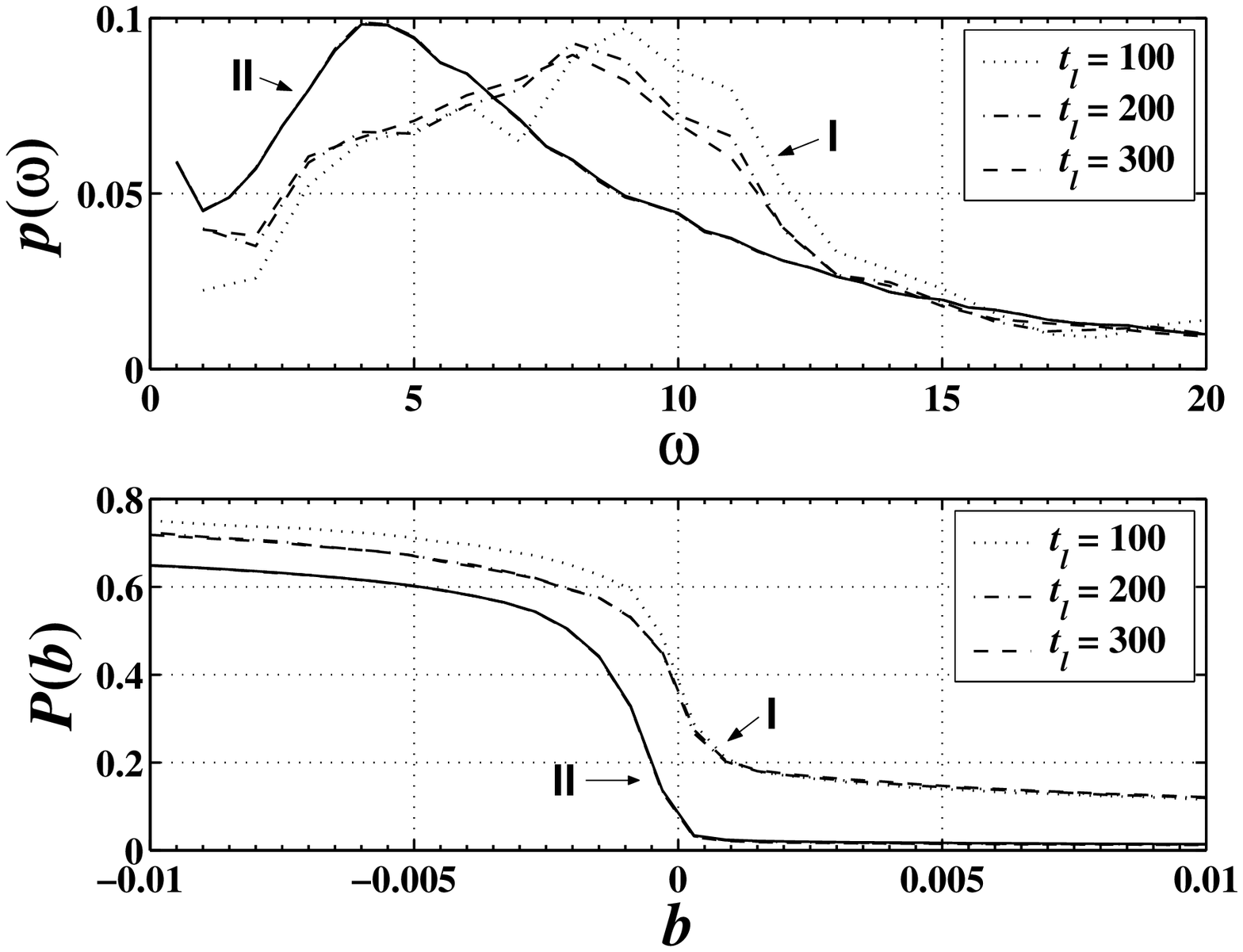}
\caption{\label{Fig1} Density distribution $p(\omega | {\rm I ~
or~ II})$ of the DSI parameter $\omega$ obtained from
Eqn.(\ref{Eq:lnpt}) and complementary cumulative distribution
$P({b}|{\rm I ~ or~ II})$ of the constraint parameter ${b}$
obtained from Eqn.(\ref{positcond}) for the objects in classes I
(dotted, dashed, and dotted-dashed) and II (continuous) for three
different values of $t_l$.}
\end{figure}

Figure \ref{Fig1} presents the distributions of the $\omega$'s
(defined in Eqn.(\ref{Eq:lnpt})) and ${b}$'s (defined in
Eqn.(\ref{defb})), obtained from fits with Eqn.(\ref{Eq:lnpt}) in
intervals ending on objects for all time scales and for three
different values of $t_l$. The distributions for classes
${\rm{I}}$ and ${\rm{II}}$, which are very robust with respect to
changes in $t_l$, are very different: for $\omega \in [6, 13]$ (which
corresponds to the values obtained in previous works),
$p(\omega | {\rm{I}})$ is larger than $p(\omega | {\rm{II}})$;
$40\%$ of objects in class ${\rm{I}}$ obey the constraint ${b}
\geq 0$ compared to $6\%$ in class ${\rm{II}}$. A standard
Kolmogorov-Smirnov test gives a probability that this difference
results from noise much less than $10^{-3}$. The very significant
differences between the distributions in classes ${\rm{I}}$ and
${\rm{II}}$ confirm that the DSI and constraint parameters are
distinguishing patterns of the three bubbles.

\begin{figure}
\includegraphics[width=13cm]{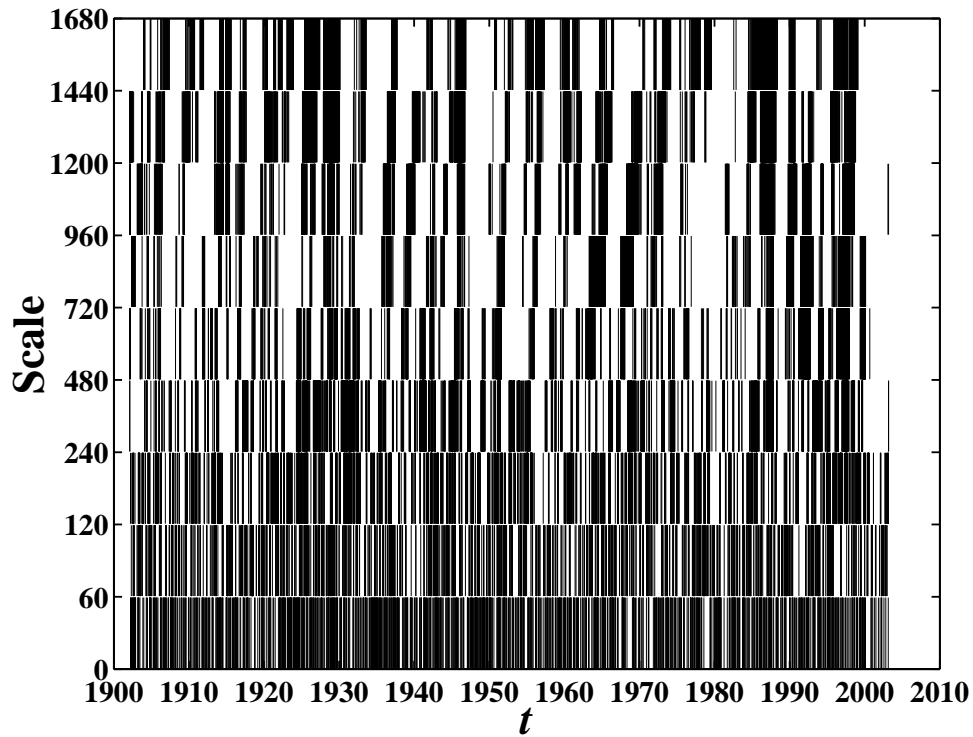}
\caption{\label{Fig2} Alarm times $t$ (or dangerous objects)
obtained by the multiscale analysis. The alarms satisfy ${b} \ge
0$, $6 \le \omega \le 13$ and $0.1 \le m \le 0.9$ simultaneously.
The ordinate is the investigation ``scale'' in trading day unit.
The results are robust with reasonable changes of these bounds. }
\end{figure}

Figure~\ref{Fig2} plots a vertical line segment for each time
scale at the date $t$ of the object that fulfills simultaneously
three criteria: $6 \le \omega \le 13$, ${b}\ge 0$ and $0.1 \le m
\le 0.9$. The last criterion ensures a smooth super-exponential
acceleration.  Each stripe thus indicates a potentially dangerous
time for the end of a bubble and for the coming of a crash. The
number of dangerous objects decreases with the increase of the
scale of analysis, forming a hierarchical structure. Notice the
clustering of alarms, especially at large scales, around the times
of the three targets. Altogether, Figs.~\ref{Fig1} and \ref{Fig2}
suggest that there is valuable information for crash prediction
but the challenge remains to decipher the relevant patterns in the
changing financial environment. For this, we call in the
statistical analysis of sparse data developed by the Russian mathematical
school \cite{Gelfand76,KBbook} in order to construct a
systematic classifier of bubbles. The modified ``CORA-3''
algorithm \cite{Gelfand76} that we briefly describe below allows
us to construct an alarm index that can then be tested
out-of-sample.

Each object is characterized by five quantities, which are
believed to carry relevant information: $\omega$, ${b}$, $m$, the
r.m.s. of the residuals of the fits with Eqn.(\ref{Eq:lnpt}) and the
height of the power spectrum $P_N$ of log-periodicity at $\omega$
\cite{Press,ZS02IJMPC}. The residuals exhibit much less structure
that the initial time series but some dependence remains as the
fits with Eqn.(\ref{Eq:lnpt}) account only for the large time-scale
structure. Indeed, using the r.m.s. of the residuals in the discrimination of
the objects is useful only if some dependence remains in them.
In other words, the residuals contain all the information at the smaller
time scales and it is hoped that these sub-structures may be linked
to the log-periodic structure at the largest time scale. This has
actually been shown to be the case by \cite{Drozdz} in a specific case
and by the construction of trading methods at short time scales
[{\it Johansen and Sornette}, unpublished]. This also follows
from the full solution of the underlying so-called renormalization group
equation embodying the critical nature of a bubble and crash
\cite{AntiBubble3}. The use of the power spectrum $P_N$ follows the same logic.

Integrated with the 9 scales, there are
totally $5\times 9$ parameters for each object. One could worry that using
so many parameters is meaningless in the deterministic framework
of Eqn.(\ref{Eq:lnpt}) because the large number of degrees of freedom
and their associated uncertainties lead to an effective randomness, but
this view does not recognize that this approach amounts
to a decomposition in multiple scales: (a) it is allowed for or recognized
that the price trajectory has patterns at many scales; (ii) for each
pre-selected scale,
a rather parsimonious deterministic fit is obtained. The selection of the
relevant parameters discussed below amounts actually to searching for
those time scales that
are the most relevant, since the data should decide and not some a priori
preconception.

We set $t_l=200$
but have checked that all results are robust with large changes of
$t_l$. Class I has 444 objects. By proceeding as in
Fig.~\ref{Fig1}, for each of the 45 parameters, we qualify it as
relevant if its distributions in classes I and II are
statistically different. We have played with several ways of
quantifying this difference and find again a strong robustness.
Out of the 45 parameters, 31 are found to be informative and we
have determined their respective intervals over which the two
distributions are significantly different. We use these 31
parameters to construct a questionnaire of 31 questions asked to
each object. The $i$th question on a given object asks if the
value of the $i$th parameter is within its qualifying interval.
Each object $A$ can then be coded as a sequence of binary
decisions so that $A=A_1A_2\cdots A_{31}$ where $A_i$ is the
answer (Y/N) to the $i$th question.

\subsection{Construction of the alarm index}

A distinctive property of the pattern recognition approach is to
strive for robustness notwithstanding the lack of sufficient
training data (here only three bubbles) \cite{Gelfand76}. In this
spirit, we define a trait as an array $(p,q,r,P,Q,R)$ where
$p=1,2,\cdots,31$, $q=p,p+1,\cdots,31$, $r=q,q+1,\cdots,31$ and
$P,Q,R=$ Y or N. There are $2 {31 \choose 1} +4 {31 \choose 2} + 8
{31 \choose 3}=37882$ possible traits. If $P=A_p$, $Q=A_q$ and
$R=A_r$, we say that the object $A$ exhibits a trait
$(p,q,r,P,Q,R)$. The number of questions (three are used here)
defining a trait is a compromise between significance and
robustness. A feature is a trait that is present relatively
frequently in class I and relatively infrequently in class II.
Specifically, if there are no less than $k_{\rm{I}}$ objects in
class ${\rm{I}}$ and no more than $k_{\rm{II}}$ objects in class
${\rm{II}}$ that exhibit a same trait, then this trait is said to
be a feature of class ${\rm{I}}$. Calling $n(\$t)$ the number of
distinctive features of a given object at time $t$, we define the
alarm index $AI(t)$, here amounting to
perform a moving average of $n(\$t)$ over a time window of width
$\Delta$. We then normalize it: $AI(t) \rightarrow
AI(t)/\max_t\{AI(t)\}$.

\subsection{Synoptic view of the approach}

Let us summarize our construction leading to the Alarm Index.
\begin{itemize}
\item We select a few targets (the three well-known speculative bubbles ending
in a crash at Oct. 1929, Oct. 1987 and Aug. 1998 for the US index
and two targets for the Hang Seng), which serve to
train our system (independently for the two indices).

\item An object is defined simply as a trading day
ending a block of trading days of a pre-defined duration.

\item Those objects which are in a neighborhood of the
crashes of the targets are defined to belong to class I. All other objects are
said to belong to class II.

\item For each object, we fit the price trajectory with expression
(\ref{Eq:lnpt}) over a given time scale (defining the duration of the block
of days ending with the object used in the fit)
and obtain the corresponding values of the parameters, which
are considered as characteristic of the object. This step is repeated
nine times, once for each of the nine time scales used in the
analysis. We keep $5$ parameters
of the fit by expression (\ref{Eq:lnpt}) for each time scale, thus
giving a total
of $5 \times 9$ parameters characterizing each object.

\item We construct the probability density functions (pdf) of each
parameter over
all objects of class I and of class II separately. Those parameters
which are found
to exhibit sufficiently distinct pdf's over the two classes are kept as being
sufficiently discriminating. In this way, out of the total of $45$ parameters,
$31$ are kept.

\item Each selected parameter gives a binary information, Y or N, as
whether its
value for a given object falls within a qualifying interval for class I.

\item As a compromise between robustness and an exhausive description,
we group the parameters in triplets (called traits) to obtain a
characterization of objects.
Ideally, one would like to use all $31$ parameters simultaneously for
each object, but the
curse of dimensionality prevents doing this.

\item We study the statistics of all traits and look for those which are
frequent in objects of class I and unfrequent in objects of class II.
Such traits
are called features.

\item The Alarm Index at a given time
is defined as a moving average number of distinctive features
found at that time. Large values of the Alarm Index are proposed to be
predictors of changes of regime in the stock market.

\end{itemize}

\subsection{Tests on the DJIA}

\begin{figure}
\includegraphics[width=13cm]{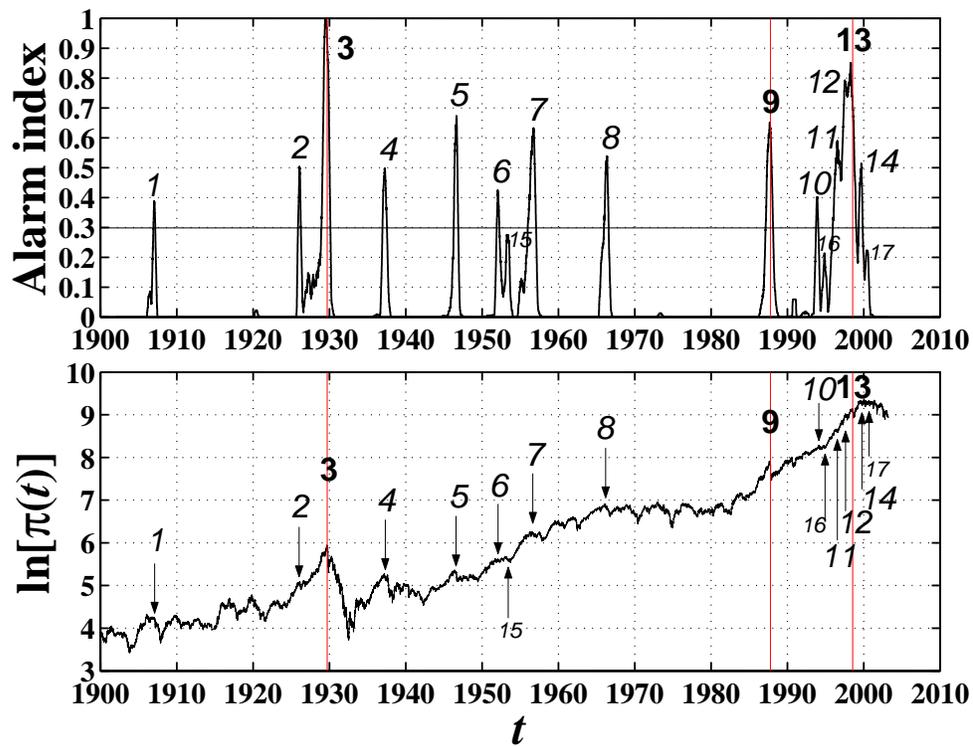}
\caption{\label{Fig3} (Color online) Alarm index $AI(t)$ (upper
panel) and the DJIA index from 1900 to 2003 (lower panel). The
peaks of the alarm index occur at times indicated by arrows in the
bottom panel. }
\end{figure}

Figure \ref{Fig3} shows the alarm index $AI(t)$ for ${\Delta}=100$
with $k_{\rm{I}}=200$ (corresponding to at least $45.1\%$ of the
objects in class I) and $k_{\rm{II}}=1500$ (corresponding to no
more than $5.6\%$ of the objects in class II). The alarm index is
found to be highly discriminative as only 13.0\% (respectively 9.5\%,
and 4.7\%)
of the total time is occupied by a high level of alarm larger than
$0.2$ (respectively $0.3$, and $0.4$). We have performed extensive
experiments to test the robustness of the results. We have varied
simultaneously $k_{\rm{I}}$ from 100 to 400 with spacing 50 and
$k_{\rm{II}}$ from $1000$ to $4000$ with spacing $500$. The
results are remarkably robust with no change of the major peaks.

The three bubbles used in the learning process are predicted (this
is expected of course): peak
3 is on 01-Jul-1929 (market maximum on 14-Sep-1929); peak 9 is on
10-Oct-1987 (market maximum on 04-Sep-1987 and crash on
19-Oct-1987); peak 13 is on 23-Apr-1998 (market maximum on
18-Jul-1998). This timing is a good compromise for a prudent
investor to exit or short the market without losing much momentum.
This success is however not a big surprise because this
corresponds to in-sample training. The most remarkable property of
our algorithm is its ability for generalizing from  the three
bubbles ending with crashes to the detection of significant
changes of regimes that may take a large variety of shapes (for
instance, a crash is not a one-day drop of, say, +15\% but can be spread
over several days or a few weeks with rather complex trajectories,
leading to very large cumulative losses).

The following peaks are predictors of crashes or of strong
corrections with an $AI=0.3$ alarm threshold: peak 1 (Jan, 1907),
peak 4 (Apr, 1937), peak 5 (Aug, 1946), peak 6 (Jan, 1952), peak 7
(Sep, 1956), peak 8 (Jan, 1966), peak 12 (Jul, 1997) and peak 14
(Sep, 1999). The more than $20\%$ price drops in 1907 (peak 1) and
1937 (peak 4) have been previously identified as crashes occurring
over less than three months \cite{MW02}. Peak 12 is occurring
three months before the turmoil on the stock market (one day $7\%$
drop on 27-Oct-1997) that was followed by a three month plateau
(see discussion in Chapter 9 of Ref.~\cite{Review1,Review2}). Peak
10 is a false alarm. Peaks 2 and 11 can be regarded as early
alarms of the following 1929 crash and 1997, 1998, and 2000
descents. Peak 15-17 are smaller peaks just barely above $AI=0.2$.
Peaks 15 and 16 are false alarms, while peak 17 (31-May-2000) is
just two months before the onset of the global anti-bubble
starting in August of 2000
\cite{AntiBubble1,AntiBubble2,AntiBubble3}.

\begin{figure}
\includegraphics[width=13cm]{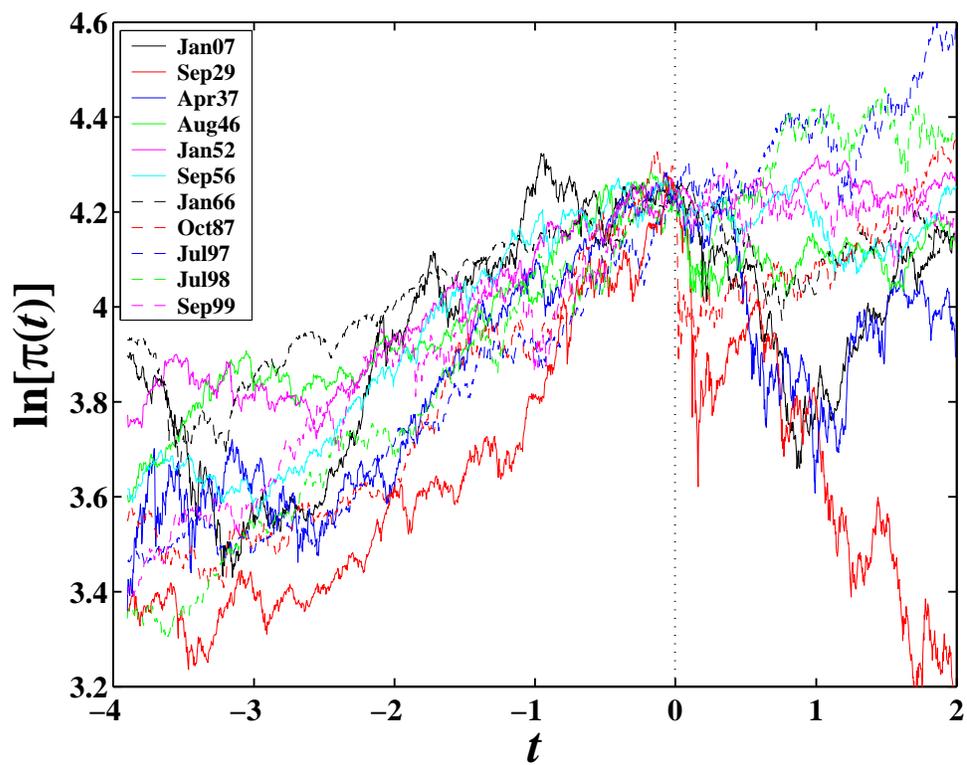}
\caption{\label{Fig4} (Color online) Superposed epoch analysis of
the 11 time intervals, each of 6 years long, of the DJIA index
centered on the time of the maxima of the 11 predictor peaks above
$AI=0.3$ of the alarm index shown in Fig.~\ref{Fig3}.}
\end{figure}

Figure~\ref{Fig4} is constructed as follows. Each alarm index peak
is characterized by the time $t_c$ of its maximum and by its
corresponding log of price $\ln[\pi(t_c)]$. We stack the index
time series by synchronizing all the $t_c$ for each peak at the
origin of time and by translating vertically the prices so that
they coincide at that time. Figure~\ref{Fig4} shows four years of
the DJIA before and two years after the time of each alarm index
maximum. We choose this representation to avoid the delicate and
still only partly resolved task of defining what is a crash, a
large drop or a significant change of regime. On this problem, the
literature is still vague and not systematic. Previous systematic
classifications \cite{dd1,dd2,MW02} miss events that are obvious
to the trained eyes of professional investors or on the contrary
may incorporate events that are controversial.

Figure~\ref{Fig4} shows that these peaks all correspond to a time
shortly before sharp drops or strong change of market regimes. In
a real-time situation, a similar prediction quality can be
obtained in identifying the peaks by waiting a few weeks for the
alarm index to drop or alternatively by declaring an alarm when
$AI(t)$ reaches above a level in the range $0.2-0.4$. The
sharpness and the large amplitudes of the peaks of the alarm index
ensures a strong robustness of this latter approach. Our method
however misses several large price drops, such as those in 1920,
1962, 1969, 1974 \cite{MW02,Review1,Review2}. This may be
attributed to a combination of the fact that some large market
drops are not due to the collective self-organization of agents
but result from strong exogenous shocks, as argued in
Refs.~\cite{endoexo,JS03a,JS03b} and that the method needs to be
improved. However, the generalizing ability of our learning
algorithm strengthens the hypothesis that herding behavior leads
to speculative bubbles and sharp changes of regimes.

\subsection{Error diagrams for evaluation of prediction performance}

\subsubsection{Theoretical background}

This brief presentation is borrowed from Chap.~9 of the book in
\cite{Review1,Review2}. In evaluating predictions and their impact
on (investment) decisions, one must weight the relative cost of
false alarms with respect to the gain resulting from correct
predictions. The Neyman-Pearson diagram, also called the decision
quality diagram, is used in optimizing decision strategies with a
single test statistic. The assumption is that samples of events or
probability density functions are available both for correct
signals (the crashes) and for the background noise (false alarms);
a suitable test statistic is then sought which optimally
distinguishes between the two. Using a given test statistic (or
discriminant function), one can introduce a cut which separates an
acceptance region (dominated by correct predictions) from a
rejection region (dominated by false alarms). The Neyman-Pearson
diagram plots contamination (misclassified events, i.e.,
classified as predictions which are thus false alarms) against
losses (misclassified signal events, i.e., classified as
background or failure-to-predict), both as fractions of the total
sample. An ideal test statistic corresponds to a diagram where the
``Acceptance of prediction'' is plotted as a function of the
``acceptance of false alarm'' in which the acceptance is close to
$1$ for the real signals, and close to $0$ for the false alarms.
Different strategies are possible: a ``liberal'' strategy favors
minimal loss (i.e., high acceptance of signal, i.e., almost no
failure to catch the real events but many false alarms), a
``conservative'' one favors minimal contamination (i.e., high
purity of signal and almost no false alarms but many possible
misses of true events).

\cite{Molchan1,Molchan2} has shown that the task of predicting an event in
continuous time can be mapped onto the Neyman-Pearson procedure.
He has introduced the ``error diagram'' which plots the rate of
failure-to-predict (the number of missed events divided by the
total number of events in the total time interval) as a function
of the rate of time alarms (the total time of alarms divided by
the total time, in other words the fraction of time we declare
that a crash or large correction is looming)
(see also \cite{KBbook} for extensions
and reviews). The best predictor corresponds to a point close to
the origin in this diagram, with almost no failure-to-predict and
with a small fraction of time declared as dangerous: in other
words, this ideal strategy misses no event and does not declare
false alarms! These considerations teach us that making a
prediction is one thing, using it is another which corresponds to
solving a control optimization problem \cite{Molchan1,Molchan2}.

\subsubsection{Assessment of the quality of predictions by the
error diagram}

To assess quantitatively the prediction procedure described above,
we thus construct an error diagram, plotting the failure to
predict as a function of the total alarm duration. The targets to
be predicted are defined as large drawdowns (cumulative losses as
defined in \cite{dd1,dd2}) whose absolute values are greater than
some given value $r_0$. Each drawdown has a duration covering a
time interval $[t_1,t_2]$. For a given alarm index threshold
$AI_0$, if there exist a time $t \in [t_1-\Delta t,t_1)$ so that
$AI(t) \ge AI_0$, then the drawdown is said to be successfully
predicted; in contrast, if $AI(t) < AI_0$ for all $t \in
[t_1-\Delta t,t_1)$, we say it is a failure. This definition
ensures that the forecast occurs before the occurrence of the
drawdown. The error diagram is obtained by varying the decision
threshold $AI_0$. The quantity ``failure to predict'' is the ratio
of the number of failures over the number of targets as said
above. The total alarm duration is the total time covered by the
objects whose alarm index is larger than $AI_0$ divided by the
total time. By varying the value of $AI_0$, we obtain different
pairs of failure to predict and total alarm duration.

\begin{figure}
\includegraphics[width=13cm]{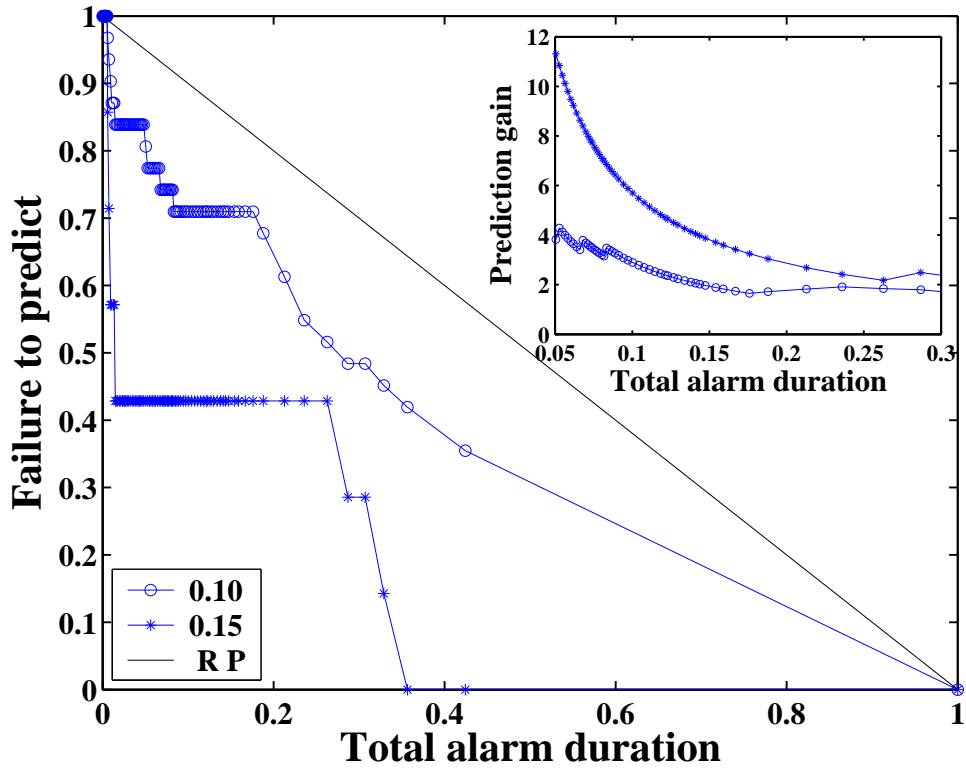}
\caption{\label{Fig5} Error diagram for our predictions for two
definitions of targets to be predicted $r_0=0.1$ and $r_0=0.15$
obtained for the DJIA. The anti-diagonal line corresponds to the
random prediction result. The inset shows the prediction gain.}
\end{figure}

Figure \ref{Fig5} presents the error diagram for two target definitions
($r_0=0.1$ and $r_0=0.15$) for the DJIA with
$\Delta t=40$, $k_{\rm{I}}= 250$ and $k_{\rm II} = 3500$. The results
do not change
significantly as long as $k_{\rm{I}}$ is much smaller than $k_{\rm II}$.
The anti-diagonal
corresponds to the completely random prediction scheme. The error diagrams with
different $\Delta t$, $k_{\rm{I}}$ and $k_{\rm II}$ are found to be similar,
with small variations, showing a strong robustness of our construction.
The inset shows the prediction gain, defined as the ratio of the fraction
of targets correctly predicted (equal to one minus the fraction of
missed targets) to the total alarm duration. Theoretically,
the prediction gain of a random prediction strategy is $1$. The error diagram
shows that
half of the targets with $r_0=0.15$ are predicted with a very
small alarm duration and all targets are predicted for an alarm
duration less than 40\%. This gives rise to large prediction gains
for small alarm durations, confirming the very good quality of our
predictors. Confidence in this conclusion is enhanced by finding that
the quality of the prediction increases as the targets are evolved
toward largest losses ($10\%$ loss for $r_0=0.1$ to $15\%$ loss for
$r_0=0.15$).
This trend continues for largest losses but there are
too few events to draw statistically meaningful conclusions for larger
drawdowns of $20\%$ or more.

\section{Tests on the Hong Kong Hang Seng composite index (HSI)}
\label{s1:HSI}

The validity of our construction should be ascertained by testing
it without modification on other independent time series. Here, we
present similar promising results obtained for the Hong Kong Hang
Seng composite index (HSI) from 24-Nov-1969 to present. The Hong
Kong Hang Seng composite index is particularly interesting as it
can be considered as a ``textbook'' example of an unending
succession of bubbles and crashes \cite{SJ01QF}. The nine biggest
crashes since 24-Nov-1969 were triggered approximately at the
following dates: 20-Sep-1971, 5-Mar-1973, 04-Sep-1978,
13-Nov-1980, 01-Oct-1987, 15-May-1989, 4-Jan-1994, 8-Aug-1997, and
28-Mar-2000. Except for the last one which was posterior to the
study published in \cite{SJ01QF}, all of them have been studied
previously in \cite{SJ01QF}. The distinctive properties of the
first eight bubbles are consistent with those reported previously
\cite{SJ01QF} (see page 461 concerning the parameters $\omega$ and
$m$). In contrast with Ref. \cite{SJ01QF} in which the positivity
constraint was not considered, we find here that it plays a
significant role in screening out solutions. We use the two
crashes in 1987 and 1997 to train the parameters of our algorithm
because they give the best defined log-periodic power law patterns
over the longest time intervals.

\begin{figure}
\includegraphics[width=13cm]{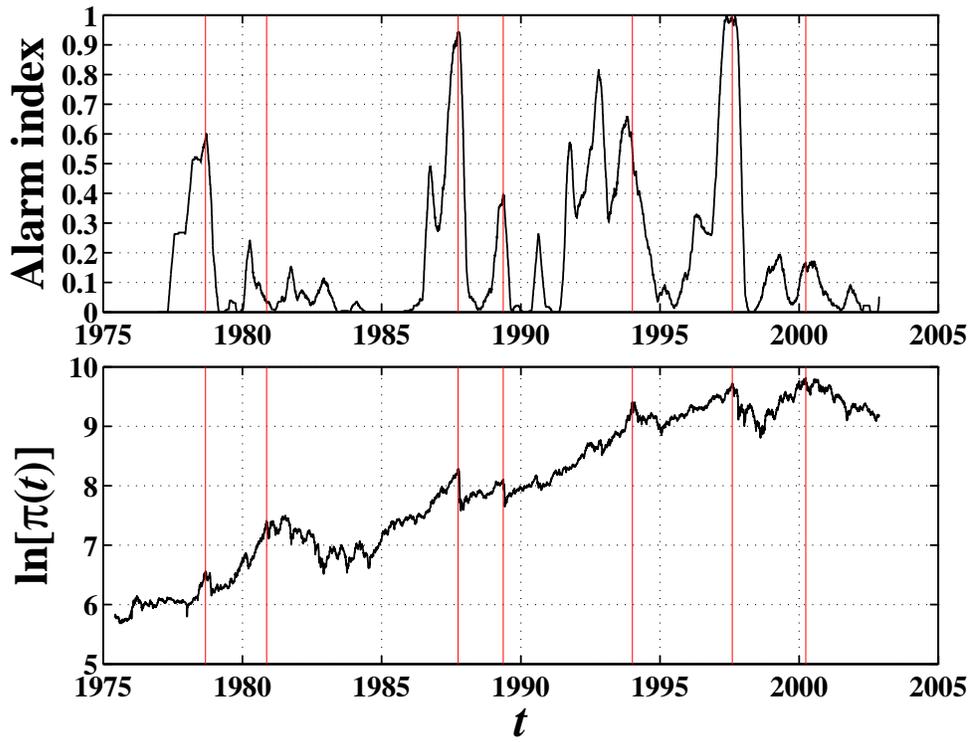}
\caption{\label{Fig6} (Color online) Alarm index $AI(t)$ (upper
panel) and the Hong Kong Hang Seng composite index from 1975 to
2003 (lower panel). The vertical lines indicate the timing of the
seven largest crashes. Note that the first two crashes are not
included in the analysis since the longest window used in our
multiscale analysis is seven years.}
\end{figure}

Figure \ref{Fig6} illustrates the typical result for the Alarm
Index obtained with $t_l=200$, $k_{\rm{I}}=210$ and $k_{\rm II}=1700$. Four
out of the five crashes that were not used in the training phase
are identified very clearly by the alarm index peaks (note that
the first two crashes in 20-Sep-1971 and 5-Mar-1973 are omitted in
our counting since they occur before the time when our Alarm Index
can be constructed). This encouraging result seems to confirm the
hypothesis that the seven crashes, posterior to 1976 where the
Alarm Index can be defined, are triggered by endogenous stock
market instabilities preceded by log-periodic power-law
speculative bubbles \cite{JS03a,JS03b}. Varying the values of the
parameters $t_l$, $k_{\rm{I}}$ and $k_{\rm II}$ of our algorithm gives
results which are robust.

In addition, one can identify more peaks in Fig.~\ref{Fig6} and it
is an interesting question to test whether they are associated
with an anomaly in the market. In total, we count 15 peaks in
Fig.~\ref{Fig6} around 15-Sep-1978 (S), 09-Apr-1980 (?),
29-Sep-1981 (F), 30-Nov-1982 (F), 01-Oct-1986 (?), 30-Sep-1987
(S), 20-May-1989 (S), 18-Aug-1990 (F), 09-Oct-1991 (F),
19-Oct-1992 (F), 28-Oct-1993 (S), 05-May-1996 (?), 06-Aug-1997
(S), 10-Apr-1999 (F), and 25-Mar-2000 (S), whose alarm indexes are
greater than $0.1$. Six peaks whose dates are identified by ``S''
correspond to successful predictions, among which two are trivial
since they were used in the training process. Six peaks identified
with ``F'' are false alarms. The remaining three peaks marked with
``?'' are neither successful predictions nor complete false alarms
as they fall close to crashes. The date with an alarm index of 0.1
on the east side of the first ``?'' alarm peak is 31-Jul-1980 and
can probably be interpreted as a forerunner of the crash on
13-Nov-1980. The two other ``?'' alarm peaks can probably
interpreted as ``fore-alarms'' of two main alarm peaks used for
training the algorithm.

\begin{figure}
\includegraphics[width=13cm]{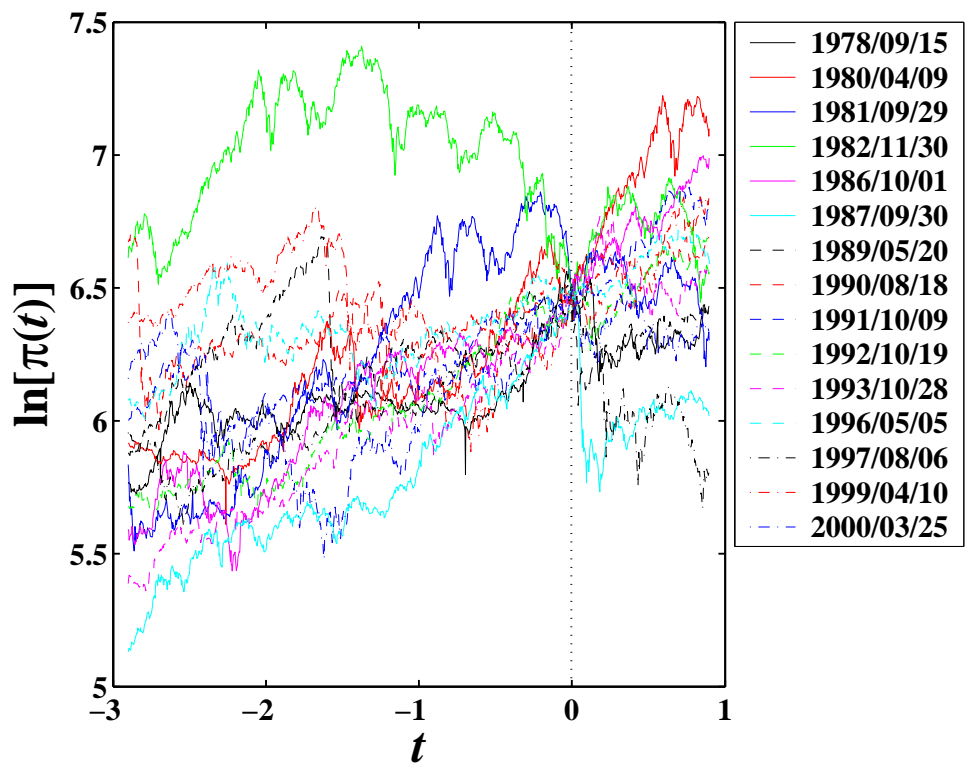}
\caption{\label{Fig7} (Color online) Superposed epoch analysis of
the 15 time intervals of the HSI index centered on the time of the
maxima of the 15 peaks above $AI=0.1$ of the alarm index shown in
Fig.~\ref{Fig6}.}
\end{figure}

Similar to Fig.~\ref{Fig4}, Fig.~\ref{Fig7} shows a superposed
epoch analysis of the price trajectories a few years before and
one year after these 15 peaks, to see whether our algorithm is
able to generalize beyond the detection of crashes to detect
changes of regimes. The ability to generalize is less obvious for
the Hang-Seng index that it was for the DJIA shown in
Fig.~\ref{Fig4}.

\begin{figure}
\includegraphics[width=13cm]{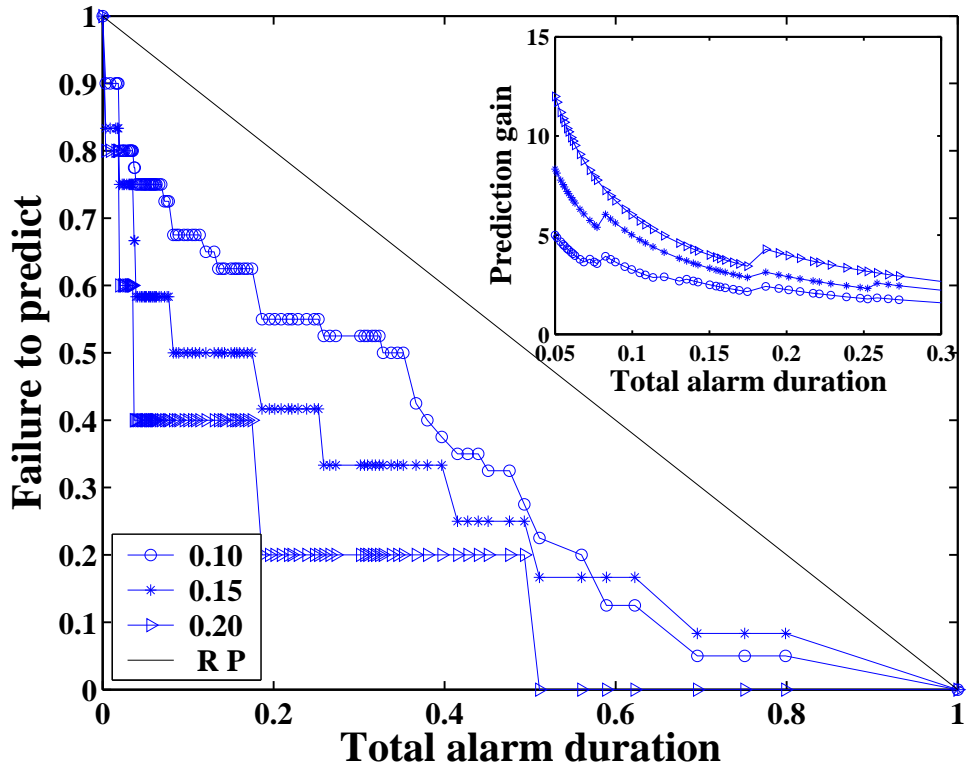}
\caption{\label{Fig8} Error diagram of the predictions for three
definitions of targets to be predicted with $r_0=0.1$, $r_0=0.15$,
and $r_0=0.2$ in regard to HSI. The diagonal line is a random
prediction. In the inset shows the prediction gains.}
\end{figure}

Figure \ref{Fig8} presents the error diagram for three target
definitions ($r_0=0.1$, $r_0=0.15$, and $r_0=0.2$) for the HSI
with $\Delta t=40$, $k_{\rm{I}}= 210$ and $k_{\rm II} = 1700$. The
anti-diagonal corresponds to the random prediction scheme. Again,
the error diagrams with different $\Delta t$, $k_{\rm{I}}$ and $k_{\rm II}$
are robust. The inset plots the corresponding prediction gains.
More than half of the targets with $r_0=0.20$ are predicted with a
very small alarm time and all targets are predicted for a finite
alarm time less than 50\%. These results are associated with a
very significant predictability and strong prediction gains for
small alarm durations.

\section{Concluding remarks}
\label{s1:conclude}

We find that different stock markets have slightly different
characteristics: for instance, the parameters trained on the DJIA
are not optimal for the Hang-Seng index. When we use the
parameters obtained from the training of our algorithm on the DJIA
on the HSI, we find that two of the seven crashes
in the HSI are identified accurately (01-Oct-1987 and 4-Jan-1994)
while the five other crashes are missed. Interestingly, these
two crashes are the most significant. This suggests the
existence of a universal behavior\footnote{
The term ``universality'' is used here in the technical sense developed
in thermodynamics and statistical physics.
Its roots go back to the 1960s when pioneers,
such as B. Widom, L. Kadanoff,  M.E. Fisher, K. Wilson and
many others explored and established the theory of critical phenomena in
natural sciences. This theory was fully developed in the 1970s to
describe the peculiar change of organization that may occur in fluids or
magnets and many other condensed matter systems. In any system in
nature, there are at least two tendencies that oppose each other:
interactions between constituents favor order while ``noise'' or
thermal fluctuations promote disorder. The referee of this fight between
order and disorder is called a ``control parameter'': by varying it, the
fluid or magnet may undergo a transition from an ordered to a disordered
state. The transition may be ``critical'' in the technical sense that
fluctuations of both competing states occur at all space and time scale
(bounded of course by the size of the system) and become intimately
intertwinned. This leads to specific signatures in the form of power law
dependences of physical observables (such as density difference or
magnetization, correlation length, susceptibility) as a function of the
distance of the control parameter to its critical value. The concept of
universality enters in this picture from the remarkable empirical
discovery later understood within the framework of the renormalization
group theory that the critical exponents of these power laws
characterizing a critical point are universal: they are the same for a
magnet or a fluid within the same ``universality class'' defined only by
very general properties of the system (such as the dimension of the
embedding space, the dimension of the order paremeter and symmetries).
The exponents are otherwise completely independent of the nature of the
system, whether it is constituted of atoms, molecules or magnetic spins.
In other words, the properties of a critical point are independent of
many of the details of a system. A major challence in the theory of
complex systems is to evaluate to what degree this concept of universality
remains valid when extended to more complex out-of-equilibrium
dynamical systems.} only for the ``purest''
event cases, while smaller events exhibit more idiosyncratic behavior.
As we have shown, training our algorithm on these two events on the HSI
improves very significantly the prediction of the other events.
This shows both a degree
of universality of our procedure and the need to adapt to the
idiosyncratic structure of each market. This situation is similar
to that documented in earthquake predictions based on pattern
recognition techniques \cite{KBbook}.

How does the performance of our system compare with those
of commonly used stochastic models of stock indices, such as GARCH?
To answer this question,
we have used a standard method to calibrate a GARCH(1,1) model
to the S\&P500. Then, the GARCH model was run to provide a prediction
of the volatility
for the next period (here taken in unit of days), but not of the sign
of the next return (which is unpredictable according to the GARCH model).
We found a quite strong predictability of the
volatility ``predicted'' by GARCH by preceding large drops (and often
large ups) of the
market but not the reverse. In other words, GARCH is not predicting
in any way future
large drops, it is the other way around: large realized drops predict
future large GARCH-predicted values of the volatility. Actually, this is
not surprising for two reasons: (i) from the structure of the GARCH model,
a previous large daily drop (or gain) immediately increases the present
amplitude of the volatility; (ii) the fact that losses predict future
increases of volatility (and not the reverse) has a rich literature
associated with the ``leverage'' effect (see for instance \cite{Fig,chanetal}
and references therein). Thus, we can conclude that the GARCH model can in
no way reproduce or even approach the detection of impending changes of regime
as we have documented here.

In summary, we have developed a multiscale analysis on the
analysis of stock market bubbles and crashes. Based on a theory of
investor imitation, we have integrated the log-periodic power-law
patterns characteristic of speculative bubbles preceding financial
crashes within a general pattern recognition approach. We have
applied our approach to two financial time series, DJIA (Dow Jones
Industrial Average) and HSI (Hang-Seng Hong Kong index). Training
our algorithm on only a few crashes in each market, we have been
able to predict the otherwise crashes not used in the training
set. Our work provides new evidence of the relevance of
log-periodic power-law patterns in the evolution of speculative
bubbles supporting the imitation theory.

We are grateful to T. Gilbert for helpful suggestions. This
work was partially supported by the James S. Mc Donnell Foundation
21st century scientist award/studying complex system.

\newpage

\end{document}